\documentclass[preprint,preprintnumbers,amsmath,amssymb]{revtex4-1}
\usepackage[dvips]{graphicx}
\usepackage{latexsym}
\usepackage{natbib}
\usepackage[usenames]{color}
\usepackage{multirow}

\definecolor{Blue}{rgb}{0,0,0.6}
\definecolor{Green}{rgb}{0,0.6,0}
\definecolor{Red}{rgb}{0.6,0,0}

\begin{document}

\title{Evolution of autocatalytic sets in a competitive percolation model}

\author{Renquan Zhang$^{1,2}$}

\affiliation{${\bf 1}$ LMIB and School of Mathematics and Systems Science,
Beihang University, Beijing, China, 100191. ${\bf 2}$ Department of Chemical and Biological Engineering, Northwestern University, Evanston, Illinois, United States of America.}


\begin{abstract}
The evolution of autocatalytic sets (ACS) is a widespread process in
biological, chemical and ecological systems which is of great
significance in many applications, such as the evolution of new
species or complex chemical organizations. In this paper, we propose
a competitive model with a $m$-selection rule in which an abrupt
emergence of a macroscopic independent ACS is observed. By numerical
simulations, we find that the maximal increase of the size grows
linearly with the system size. We analytically derive the threshold
$t_{\alpha}$ where the abrupt jump happens and verify it by
simulations. Moreover, our analysis explains how this giant
independent ACS grows and reveals that, as the selection rule
becomes more strict, the phase transition is dramatically postponed,
and the number of the largest independent ACSs coexisting in the
system increases accordingly. Our research work deepens the
understanding of the evolution of ACS and should provide useful
information for designing strategies to control the emergence of ACS
in corresponding applications.
\end{abstract}

\maketitle

\section{Introduction}

Complex networks \cite{Albert2002} are often used to describe a
variety of chemical, biological and social systems. For instance,
the metabolism of a cell is a network of substrates and enzymes
interacting via chemical reactions \cite{Fell2000,Jeong2000}.
Ecosystems are networks of biological organisms with predator-prey,
competitive or symbiotic interactions \cite{Williams2000,
Camacho2002}. In real-world systems, these networks are by no means
static. On the contrary, biological or chemical networks often
evolve into certain structure to optimize functionality according to
some mechanisms. It is shown that the small-world structure of
metabolic networks may have evolved to enable a cell to react
rapidly to perturbations \cite{Fell2000}. Similarly, the visual
cortex may have evolved into a small-world architecture since it
would aid the synchronization of neuron firing patterns
\cite{Watts1998}. Therefore, understanding the mechanisms
responsible for the evolution of such networks is an important issue
in many applications.

To explore the mechanisms underlying the network evolution, a set of
models based on artificial chemistry of catalyzed reactions are
proposed \cite{Kauffman1986,Farmer1986,Fontana1994,Stadler1994}. An
artificial chemistry is a system whose components react with each
other in a way analogous to molecules participating in chemical
reactions. This kinds of systems are widespread in biological
research, such as the protein or enzymes within a cell
\cite{Fell2000,Jeong2000}, or some organic molecules in a pool on
the prebiotic Earth \cite{Joyce1989}. The networks evolve over time
as mutation happens \cite{Bak1993,Jain1998,Jain2002}, and the
structure of networks will in turn affect the subsequent evolutions.
With these models, questions about self-organization, the origin of
life and other evolvability issues are explored in works on
artificial chemistries
\cite{Kauffman1986,Farmer1986,Fontana1994,Stadler1994}.

Based on this framework, Bak and Sneppen \cite{Bak1993} introduced a
simple and robust model of biological evolution of an ecology of
interacting species, which had the feature that the least fit
species mutated. The model self-organized into a critical steady
state with intermittent coevolutionary avalanches of all sizes
\cite{Bak1993}. Later on, inspired by the Bak-Sneppen model, Jain
and Krishna \cite{Jain1998,Jain2002} proposed a similar model, in
which the mutation of a species also changed its links to other
species. They investigated how the network of interactions among the
species evolved over a longer time scale and the growth of the
autocatalytic set (ACS) \cite{Jain1998}. It was shown that, starting
from a sparse random graph, an ACS inevitably appeared and triggered
a cascade of exponentially increasing connectivity until it spanned
the whole graph.

The concept of an ACS is introduced in the context of a set of
catalytically interacting molecules. It is defined to be a set of
molecular species that contains, within itself, a catalyst for each
of its member species \cite{Eigen1971,Kauffman1971,Rossler1971}.
Mathematically, in a graph of interacting agents, an ACS is defined
as a subgraph whose every node has at least one incoming link from a
node that belongs to the same subgraph \cite{Jain1998}. This
definition is meant to capture the property that an ACS has
"catalytic closure" \cite{Kauffman1986}, i.e., it contains the
catalysts for all its members. Therefore, autocatalytic sets might
be more stable to perturbations because of their ability to
self-replicate. On the prebiotic Earth, autocatalytic sets are
suggested as one of the possible means by which a complex chemical
organization could have evolved \cite{Jain2001}. Due to its property
of self-replicating, the ACS plays an important role in the overall
dynamics in chemical or biological networks. As defined, an ACS may
have several disconnected components. These components are
independent units that posses property of self-replicating. We
define each component as an independent ACS and focus on the largest
one in this paper, which contains the largest number of species.

The definition of the largest independent ACS is somewhat analogous
to that of the giant component (GC) in percolation transition
\cite{Stauffer1994}. Percolation can be interpreted as the formation
of a giant component in networks. One important model to show this
process is the classic Erd\'{o}s and R\'{e}nyi (ER) \cite{Erdos1960}
model. In ER model, the evolution proceeds as follows: Starting with
$N$ isolated nodes, an edge is connected between a randomly selected
unconnected pair of nodes at each time step. Then as the number of
connected edges increases, a macroscopic cluster, i.e., the giant
component, appears at the percolation threshold, and its size grows
continuously. Recently, based on the ER model, an explosive
percolation (EP) \cite{Achlioptas2009} model was introduced. In this
model, the ER model was modified by imposing additionally a
so-called product rule or sum rule, which suppresses the formation
of a large cluster \cite{Achlioptas2009}. Because of this
suppressive bias, the percolation threshold is delayed. When the
giant component eventually emerges, it does so explosively. This
result has attracted much interest
\cite{Ziff2009,Radicchi2009,Araujo2010,Cho2013,Chen2011,Zhang2013,Costa2010,Riordan2011,Grassberger2011,Lee2011,Bastas2011}.
Initially, this explosive phase transition was regarded as a
discontinuous transition. However, it was recently found that the
transition is continuous in the thermodynamic limit
\cite{Costa2010}, followed by a mathematical proof
\cite{Riordan2011} and extensive supporting simulations
\cite{Grassberger2011,Lee2011,Bastas2011}.

Inspired by the EP model, we propose an evolving network model that
presents an abrupt emergence of the largest independent ACS. By
imposing a selection rule \cite{Cho2013,Chen2011,Zhang2013}, the
formation of ACS's largest component is suppressed. In the next
section, we introduce the definition of this competitive model, and
then discuss its properties. After that, we give an analytical
analysis of the threshold where a macroscopic independent ACS
appears. The theoretical results are verified by simulations. In the
last section, we conclude our findings and give a brief discussion.

\section{Model}

The system is described by a directed graph, in which the $N$ nodes
represent the species or chemicals and the directed links stand for
catalytic interactions between them. The graph can be completely
described by an adjacency matrix $C=\{c_{ij}\}_{N\times N}$, where
$c_{ij}=1$ if there exists a link from node $j$ to $i$, and zero
otherwise. A directed link from node $j$ to $i$ means that $i$ is
catalyzed by $j$. Specifically, we exclude the self-replicating
species, i.e. $c_{ii}=0$ for all $i=1,2,\cdots,N$.

\begin{figure}
\centering \resizebox{1\columnwidth}{!}{
\includegraphics{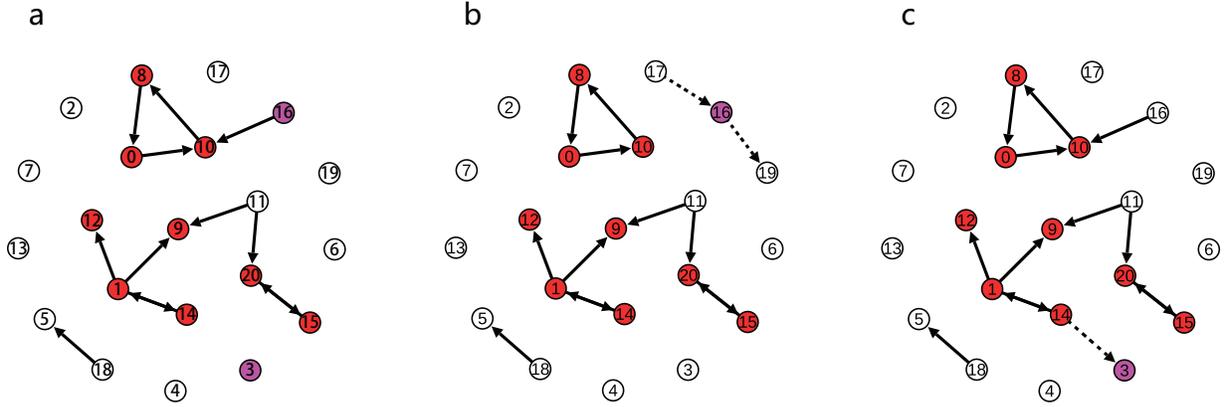}
}
\caption{{\bf Schematic illustrations of network evolution.} ($a$)
It shows the structure of the graph at some time step. There are
three independent ACSs (marked with red color) with size 2, 3, and 4
respectively. The pink nodes represent the nodes that are randomly
selected from the {\it least fitness}. ($b$) Update process for node
16. Remove the link $\{16\to 10\}$ and replace it by two randomly
selected links. ($c$) Update process for node 3. After this update,
node 3 will become a member of an independent ACS.}
\label{fig:1}       
\end{figure}

According to the definition, an ACS may consist of several disjoint
smaller ACSs, which have no intersections with each other and form
as independent catalytic systems. To describe this fact, we
introduce the concept of independent ACS. Concretely, for the nodes
in ACS, an independent ACS is the maximal weakly-connected subgraph
which is only composed of ACS-nodes. For example, in
Fig.\ref{fig:1}(a), there are three independent ACSs marked with red
color. Notice that, node $11$ and $16$ do not belong to the
independent ACS since they are not ACS-nodes, even though they have
links connecting to independent ACSs. Denote the size of independent
ACSs as $S_1, S_2, \cdots$ in descending order. Particularly, since
there may be several independent ACSs with the same size $S_i$, we
denote them as $S_{i,1}, S_{i,2}, \cdots, S_{i,{n_i}}$. Here $n_i$
represents the number of independent ACS with size $S_i$. For
convenience, we simply denote $S_i^{'}$ as the set of the
independent ACSs with size $S_i$. Based on the definition of ACS,
$S_i^{'}$ is still an ACS with $n_i S_i$ nodes. In this paper, we
focus on the largest independent ACSs ($S_1^{'}$), which usually
play the most important role of self-replicating in the evolution
process.

At the beginning time $t=0$, the initial graph is a random graph
with average indegree (or outdegree) $d$ (i.e. the probability of
linking a directed edge is $p=d/N$). For every discrete time step
$t=1,2,\cdots$, we denote $n_i(t)$ as the number of the independent
ACSs in $S_i^{'}$ at time $t$. The graph is updated as follows:

{\bf Step 1.} Select $m$ "most mutating" nodes in the network. This
procedure is performed based on a special set named {\it least
fitness} \cite{Jain1998}, which will be explained later.

{\bf Step 2.} For each of the selected nodes $i$, determine the size
of the largest independent ACS in the new network if node $i$ is
updated. The update associated with node $i$ processes as follows:
for node $i$, remove all the incoming and outgoing links attached to
it, and then replace them by links randomly connected to other nodes
with the same probability $p$. So for each selected node $i$, we get
a new adjacency matrix $C(i)$ and a new largest size of the
independent ACS $S_1(i)$.

{\bf Step 3.} Randomly select a node $i$ with $S_1(i) \leq S_1$,
where $S_1$ is the size of the largest independent ACS before the
update. If all the $S_1(i)$ are larger than original $S_1$, we just
choose one node $i$ uniformly from the $m$ selected nodes. Then the
network is updated to $C(i)$.

At each time step, we need to select $m$ nodes with "least fitness".
To achieve this, each node $i$ is assigned a population $y_i\geq 0$
and a relative population $x_i=y_i/Y$, where $Y=\sum_i y_i$. Between
two successive graph updates, the evolution of population is given
by
\begin{equation}
\frac{\mathrm{d}y_i}{\mathrm{d}t}=\sum_{j=1}^N c_{ij}y_j-\phi y_i.
\end{equation}
From the above equation, $x_i$ has the dynamics
\begin{equation}
\frac{\mathrm{d} x_{i}}{\mathrm{d}t}=\sum_{j=1}^{N}
c_{ij}x_{j}-x_{i}\sum_{k,j=1}^{N} c_{kj} x_{j}.
\end{equation}
In this model, we can use $x_{i}$ to measure the fitness of the
species $i$ in the environment defined by the graph. The larger
$x_{i}$ is, the more fit species $i$ is. So the $m$ nodes with
smallest values of $x_{i}$ are picked as the "most mutating" nodes.
Moreover, when $x_{i}$ reaches to the stable solution, the value of
$x_{i}$ is zero for almost all the nodes outside the ACS. Therefore,
before ACS spans across the whole graph, the selected nodes are
almost outside the ACS.

Fig.\ref{fig:1} illustrates the rule of the model with $m=2$. At
some step, two nodes $16$ and $3$ are selected from the set {\it
least fitness} and one of them will be updated. In the case of
Fig.\ref{fig:1}(b), the update of node $11$ will not affect the size
of the largest independent ACS. Whereas, in Fig.\ref{fig:1}(c), node
$3$ will join in the largest independent ACS, making its size
increase by 1. According to the rule, node $11$ will be picked as
the updating node since it will not increase the size of the largest
independent ACS.

When $m=1$, our model degenerates to the classic ACS evolutionary
model \cite{Jain1998}, in which the size of ACS grows exponentially.
However, if $m$ is larger than $1$, the evolution process will
present essential difference. Specifically, the size of the largest
independent ACS $S_1$ will "jump" by the size of $O(N)$ at one
single step. This discontinuous increase stems from the coexistence
of several independent ACSs with maximal size during the evolution.
We are interest in the number of the largest independent ACSs $n_1$
that the system can maintain in the process. In the next section, we
will discuss this phenomenon in more details.

\section{Properties}

\begin{figure}
\centering \resizebox{1\columnwidth}{!}{
\includegraphics{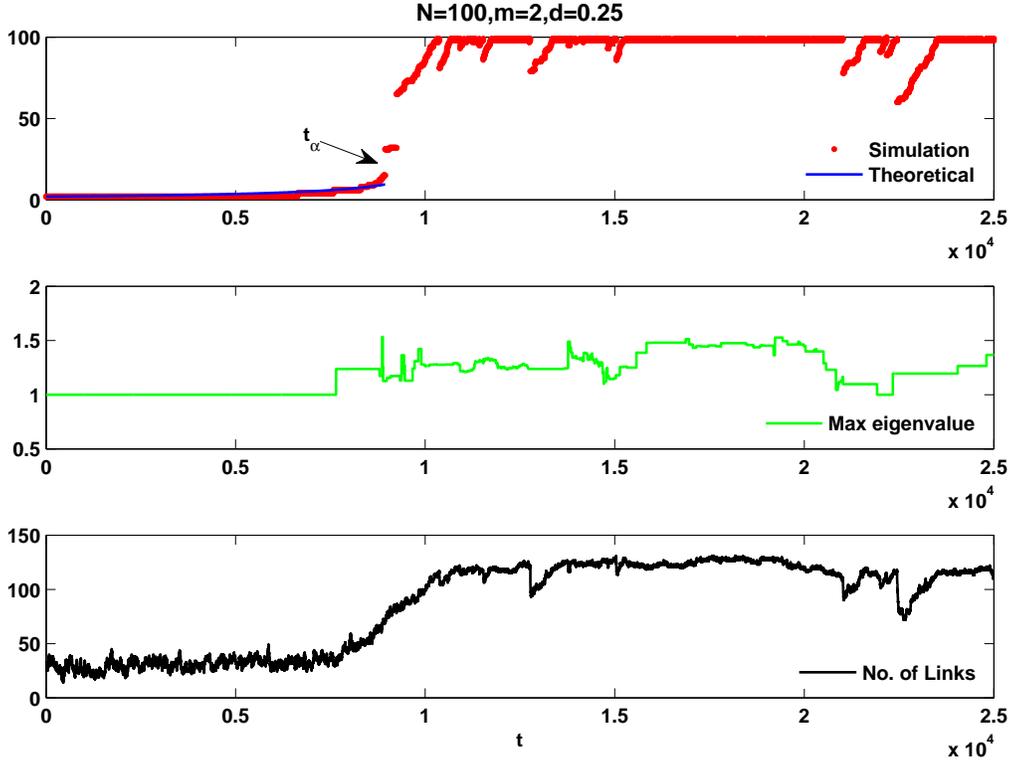}
}
\caption{{\bf The evolution process of the model with N=100, m=2 and
d=0.25.} ($a$) It shows the size of the largest independent ACS
$S_1$ after the first ACS appears. At the threshold $t_\alpha$,
$S_1$ jumps by the size of over $0.1N$. The blue line shows the
theoretical result from $t_0$ to $t_\alpha$ (see Eq.8 in section 4). ($b$) The maximal
eigenvalue of the adjacency matrix $C(t)$ versus time. During the
process, it keeps larger than $1$. ($c$) The number of links in the
system versus time. The tendency is same with the size of the
largest independent ACS.}
\label{fig:2}       
\end{figure}

Since the evolution process of the model with $m=1$ has been
investigated in \cite{Jain1998}, we focus on the situation where $m$
is larger than $1$. Fig.\ref{fig:2} shows the evolution process of
the model with $N=100$, $m=2$ and $d=0.25$. Before the first ACS
appears, almost all the nodes are in the least fitness set. Based on
this fact, every graph can be roughly assumed as a sample of the
classical ER directed random networks. In the evolution, usually the
first ACS is generated in the form of a cycle, which is also the
simplest pattern of ACS. When we update a single node, according our
assumption, a cycle will appear with the probability
$q\equiv\sum_{i=0}^{\infty} p^{i+2}N^{i+1}=\frac{Np^2}{1-Np}$, where
$p^{i+2}N^{i+1}$ is the probability of forming a cycle of size
$i+2$. As we select $m$ nodes at every step, according to the
selection rule, the probability of forming a cycle will be $q^m$. If
taking $t_0$ as the time that the first ACS appears, the
distribution of $t_0$ can be approximated by a geometric
distribution with $p(t_0=k)=q^m(1-q^m)^{k-1}$, and the expectation
of $t_0$ is $1/q^m$. Although $t_0$ is very large, the appearance of
an ACS is inevitable. For convenience, take $t_0$ as the beginning
time $0$ in Fig.\ref{fig:2}. Before $t_0$, both $S_1$ and maximal
eigenvalue are always zero, and the number of links fluctuates
around the expectation value $dN=25$.

In Fig.\ref{fig:2}(a), an abrupt jump of $S_1$ is observed at time
$t_{\alpha}$. This is different from the continuous growth of ACS in
the classical model. We will explain this in more details later. In
Fig.\ref{fig:2}(b), we present the evolution of the largest
eigenvalue of $C$. Based on the graph theory, we can get following
conclusions \cite{Jain1998}: (1) An ACS always contains a cycle. (2)
If a graph has no ACS, then the largest eigenvalue is zero. (3) If a
graph has an ACS, then the largest eigenvalue is larger than $1$.
There is no ACS before $t_0$, so the largest eigenvalue keeps zero
in this period. As shown in Fig.\ref{fig:2}(c), the number of links
changes dramatically when $S_1$ grows to the whole graph or
collapses down in Fig.\ref{fig:2}(a). Both the curves in
Fig.\ref{fig:2}(a) and (c) have the same tendency, and they also
have a strong relationship with the {\it least fitness} set.

\begin{figure}
\centering \resizebox{1\columnwidth}{!}{
\includegraphics{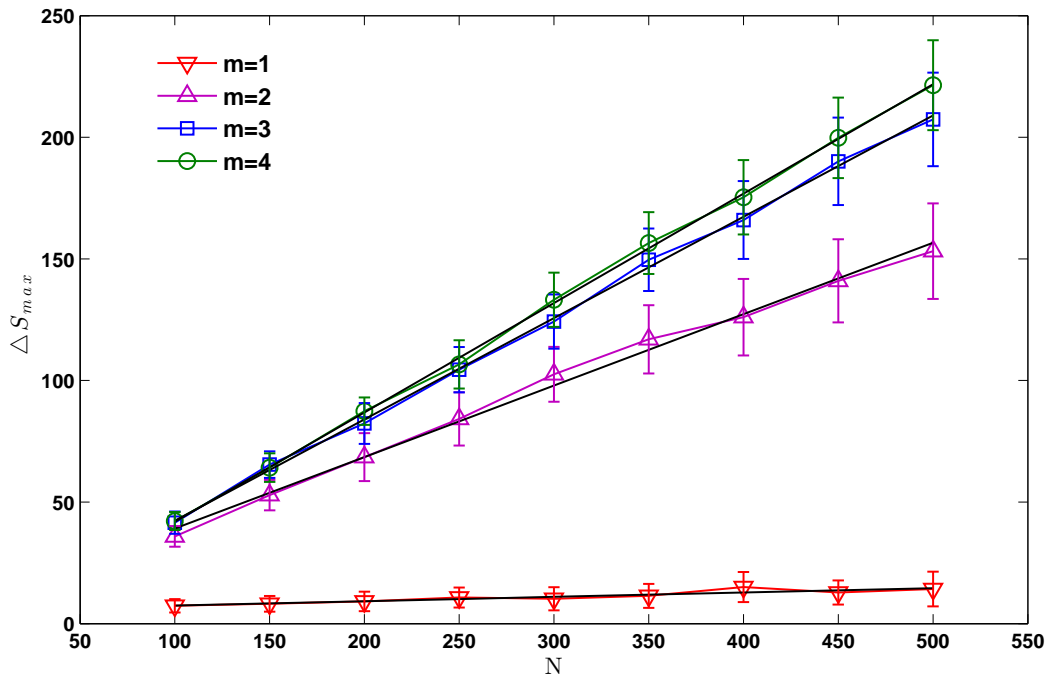}
}
\caption{{\bf Maximal jump of the $S_1$.} $\triangle S_{max}$
represents the maximal increase of $S_1$ for an update of a single
node before the independent ACS spanning across the whole graph. The
data points are averaged on 100 random instances and the error bars
are the standard deviations. It shows that the $\triangle S_{max}$
is linear with the system size $N$ for $m=2, 3$ and $4$. The slopes
of the fitting lines (black lines) are 0.018, 0.2938, 0.4166 and
0.4498 for $m=1, 2, 3$ and $4$ respectively.}
\label{fig:3}       
\end{figure}

The nature of the discontinuous jump of $S_1$ is better revealed in
Fig.\ref{fig:3}. The maximal increase of the largest independent ACS
$\triangle S_{max}$ is linear with the system size $N$ for $m\geq
2$. With this trend, we have $\lim_{N \to \infty} \frac{\triangle
S_{max}}{N}>0$. This evidence from simulation shows that our model
can lead to an abrupt emergence of a macroscopic independent ACS. In
the evolution process, there are several large independent ACSs
coexisting and finally merging together to overtake the previous
largest one. With the increase of $m$, it is more difficult for
$S_1$ to grow. As the suppression strengthens, the system can
maintain more largest independent ACSs, and there will be less nodes
outside the ACS. Moreover, it is almost impossible to break up a
large ACS into smaller ones since the updating nodes mostly don't
belongs to ACSs. For large $m$, the increase of $S_1$ will mainly
depend on merging two giant independent ACSs rather than adding
isolated nodes. In the next section, we will analyze this process
theoretically and perform simulations to verify the results.

\section{Theoretical analysis}

Based on the investigation above, it is clear that the models with
$m=1$ and $m\geq2$ are very different. In \cite{Jain1998,Jain2001},
the properties of the model with $m=1$ have been discussed with
analytical method, so in this section we will analyze the model with
$m\geq2$. First, we consider the probability of adding a node to the
ACS that has $S$ nodes (donated as $P_{add}$). Based on the
definition of ACS, a node could be added to ACS if and only if there
is an incoming link from the ACS to this node. So the probability
$P_{add}$ is:
\begin{equation}
P_{add}=1-(1-p)^{S}.\label{Padd}
\end{equation}
Here $p=d/N$ is the probability of linking an edge in update. As our
model is applied to sparse graphs, in which the linking probability
$p$ is very small, Eq.\ref{Padd} can be approximated by
\begin{equation}
P_{add}\approx pS. \label{ps}
\end{equation}
At each time step, we only update one node out of the $m$ selected
nodes. The size $S_1$ would change if all the $m$ nodes had at least
one incoming link from $S_1^{'}$. The size of $S_1^{'}$ is
$n_1(t)S_1$ and based on Eq.\ref{ps}, we can calculate the average
change of $S_1$ for the sparse graph
\begin{equation}
\triangle S_1=(n_1(t) p S_1)^m\triangle t. \label{S1}
\end{equation}
If we take the first time that an independent ACS appears as the
beginning step $t_0$, Eq.\ref{S1} can be integrated from $t_0$ to
$t_1$
\begin{equation}
\int_{S_1(t_0)}^{S_1(t_1)} \frac{1}{S_1^m}\mathrm{d}
S_1=p^m\int_{t_0}^{t_1} n_1^m(t)\mathrm{d}t.
\end{equation}
Taking $S_0$ as the initial size of $S_1(t_0)$, the equation above
can be written as follows
\begin{equation}
\frac{1}{S_0^{m-1}}-\frac{1}{S_1(t)^{m-1}}= (m-1)p^m\int_{t_0}^{t_1}
n_1^m(t)\mathrm{d}t. \label{evolution}
\end{equation}
For the last part of the Eq.\ref{evolution}, we define a quantity
$n_\beta$ such that $\int_{t_0}^{t_1}
n_1^m(t)\mathrm{d}t=n_\beta^m(t_1-t_0)$. In fact, $n_\beta$ can be
viewed as the average tolerance of the largest independent ACSs
during the evolution. In other words, the larger $n_\beta$ is, the
more largest independent ACSs the system can maintain. With this
definition, we get the function of $S_1(t)$ for different values of
$m$
\begin{equation}
S_1(t)={\frac{1}{1/S_0^{m-1}-n_\beta^m(m-1)p^m(t-t_0)}}^{\frac{1}{m-1}}.\label{s1t}
\end{equation}
The evolution function for $m=1$ should be $S_1(t)\sim
S_1(t_0)e^{p(t-t_0)}$, which presents an exponential and continuous
increase of the ACS size. This is quite different from the function
of $m\geq 2$ above. For the parameter $m\geq 2$, when $t$ takes
certain value, the denominator in Eq.\ref{s1t} will become zero.
Therefore, we identify a phase transition at some time step $t$. The
threshold $t_{\alpha}$ is
\begin{equation}
t_{\alpha}-t_0=\frac{1}{n_\beta^m(m-1)p^mS_0^{m-1}},
\end{equation}
or
\begin{equation}
\frac{t_{\alpha}-t_0}{N^m}=\frac{1}{n_\beta^m(m-1)d^mS_0^{m-1}}.\label{threshold}
\end{equation}

\begin{figure}
\centering \resizebox{1\columnwidth}{!}{
\includegraphics{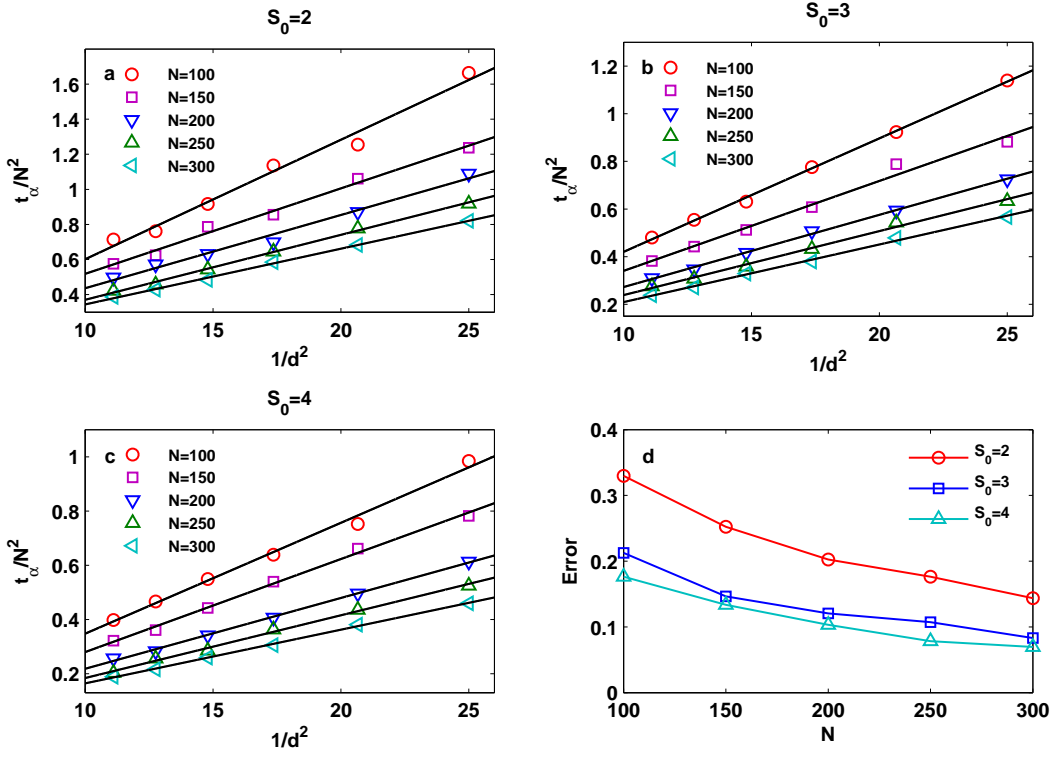}
}
\caption{{\bf The threshold $t_\alpha$ with $m=2$ versus average
degree $d$ for different $S_0$.} Different colors represent distinct
system sizes. The data are obtained by averaging 100 instances in
each case. The black lines are the linear fits with least-quare
regression. ($a-c$) Results with different initial size $S_0$ from
2, 3 and 4. ($d$) Errors for different system size $N$. Error is
defined as the average distance between the data points and the
theoretical line.}
\label{fig:4}       
\end{figure}

\begin{figure}
\centering \resizebox{1\columnwidth}{!}{
\includegraphics{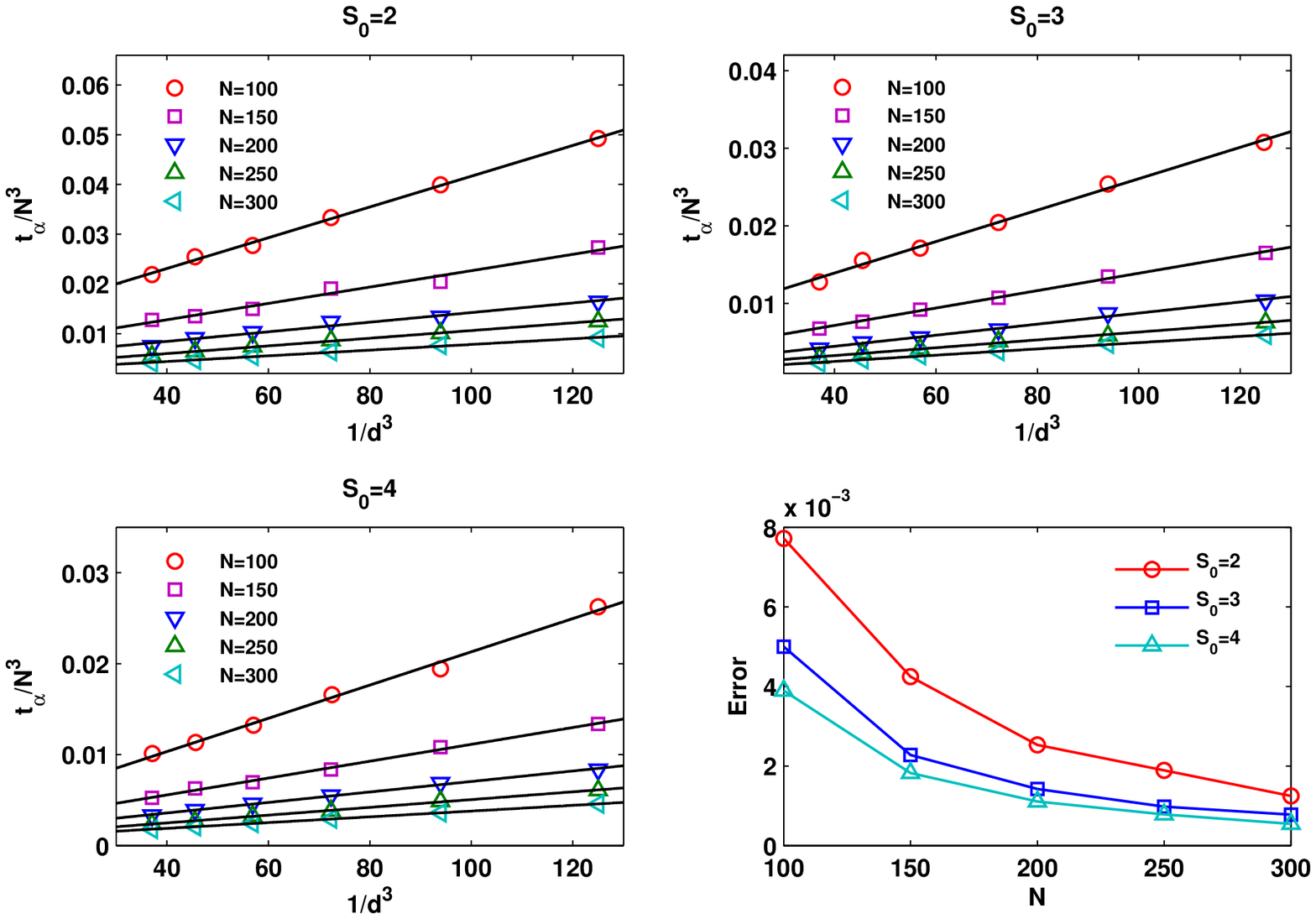}
}
\caption{{\bf The threshold $t_\alpha$ with $m=3$ versus average
degree $d$ for different $S_0$.} Different sizes are marked with
distinct colors and symbols. Each data point is obtained by
averaging 100 simulations. The black lines are the fitting lines.
($a-c$) Results with different initial size $S_0$ from 2,3 and 4.
($d$) Fitting errors versus $N$.}
\label{fig:5}       
\end{figure}

Fig.\ref{fig:4} and Fig.\ref{fig:5} are the numerical results with
$m=2$ and $m=3$ respectively. Because of the assumption of the
sparse graph, we choose the average indegree $d$ from $0.2$ to
$0.3$. For fixed value of $S_0$, we perform the simulations starting
from an initial ACS with size $S_0$ on random graphs. To determine
the threshold, in simulations we take the time step where the
increase of the largest independent ACS's size $S_1$ exceeds $10\%$
of the system size $N$ as $t_{\alpha}$. Since the starting time
$t_0$ is set to $0$, we rearrange Eq.\ref{threshold} as follows
\begin{equation}
\frac{t_{\alpha}}{N^m}=(\frac{1}{n_\beta^m(m-1)S_0^{m-1}})\cdot\frac{1}{d^m}.\label{threshold1}
\end{equation}
Therefore, there is a linear relation between the quantity
$t_{\alpha}/N^m$ and $1/d^m$. In Fig.\ref{fig:4} and
Fig.\ref{fig:5}, there is clearly a linear relation between these
two quantities, and the numerical results agree with the fitting
line very well. To check the fitting errors for different system
sizes, we define the fitting error as the average distance from the
data points to corresponding fitting line. For both cases, the
fitting errors decrease as the system size increases.

Moreover, from the slope (donated as $k$) of the fitting line we can
obtain $n_{\beta}$ by relation
\begin{equation}
k=\frac{1}{(m-1)S_0^{m-1}n_\beta^m}.
\end{equation}
In order to compare with the real number of the largest independent
ACSs coexisting in the system during the evolution, we record this
number in each time step in simulations. Then we take the average of
these values as the real $n_{\beta}$. Table.\ref{table1} and
Table.\ref{table2} display the results of $n_{\beta}$ from both the
theoretical analysis and simulations. As $m$ increases from $2$ to
$3$, $n_{\beta}$ grows a lot, which means more strict suppression
will make the system maintain more largest independent ACSs during
the evolution process. This fact will further lead to the result
that the jump of $S_1$ enhances a lot when $m$ increases, as shown
in Fig.\ref{fig:3}. Besides, as the system size $N$ grows,
$n_{\beta}$ also increases slightly. This is a natural result of the
increase of system size. Therefore, the choice of $m$ will
dramatically affect the formation of the giant independent ACS, both
the emergence time and the jump of size.

\begin{table}[h]

\begin{tabular}{|c||c||c||c||c||c|}

\hline
 & $N=100$ & $N=150$ & $N=200$ & $N=250$ & $N=300$\\
\hline
$S_0=2$ & 2.7092 & 3.2034 & 3.4603 & 3.6747 & 3.9670 \\
 & 2.7234 & 3.1138 & 3.4057 & 3.6424 & 3.8506 \\
 & 0.52\% & 2.88\% & 1.60\% & 0.89\% & 3.02\% \\
\hline
$S_0=3$ & 2.6455 & 2.9722 & 3.3169 & 3.5234 & 3.7140 \\
 & 2.4982 & 2.8377 & 3.1036 & 3.3963 & 3.5553 \\
 & 5.90\% & 4.74\% & 6.87\% & 3.74\% & 4.46\% \\
\hline
$S_0=4$ & 2.4724 & 2.6988 & 3.0940 & 3.2862 & 3.5529 \\
 & 2.2683 & 2.5943 & 2.8485 & 3.0770 & 3.3001 \\
 & 8.99\% & 4.03\% & 8.62\% & 6.80\% & 7.66\% \\
\hline
\end{tabular}

\centering \caption{{\bf The comparison of $n_\beta$ between the
simulations and the theoretical results in Fig.4(a)-(c) with $m=2$.}
The first line of every group shows the theoretical results and the
second line represents results from simulations. The third line is
the relative errors of theoretical results to real values. Each
simulation result is obtained by averaging values from 100
simulations.} \label{table1}

\end{table}

\begin{table}[h]

\begin{tabular}{|c||c||c||c||c||c|}

\hline
 & $N=100$ & $N=150$ & $N=200$ & $N=250$ & $N=300$\\
\hline
$S_0=2$ & 7.3929 & 9.1257 & 10.8941 & 11.7463 & 12.9772 \\
 & 7.2431 & 8.7351 & 10.3443 & 11.3490 & 12.4997 \\
 & 2.07\% & 4.47\% & 5.31\% & 3.50\% & 3.82\% \\
\hline
$S_0=3$ & 6.4972 & 7.9128 & 9.2004 & 10.3262 & 11.1204 \\
 & 6.1639 & 7.4202 & 8.6414 & 9.8103 & 10.5035 \\
 & 5.41\% & 6.64\% & 6.47\% & 5.26\% & 5.87\% \\
\hline
$S_0=4$ & 5.5494 & 6.9597 & 8.1509 & 9.0114 & 9.9579 \\
 & 5.2646 & 6.4332 & 7.4965 & 8.3728 & 9.2047 \\
 & 5.41\% & 8.18\% & 8.73\% & 7.63\% & 8.18\% \\
 \hline
\end{tabular}

\centering \caption{{\bf The comparison of $n_\beta$ between the
simulations and the theoretical results in Fig.5(a)-(c) with $m=3$.}
The first line of every group shows the theoretical results and the
second line represents results from simulations. The third line is
the relative errors of theoretical results to real values. Each
simulation result is obtained by averaging values from 100
simulations.} \label{table2}

\end{table}

\section{Conclusions}

ACS is an important concept in the evolution dynamics of biological,
chemical and ecological systems. The emergence of an ACS is often
used to explain the mechanism by which a complex chemical
organization or species could have evolved. In this paper, by
imposing a $m$-selection rule, we propose a competitive model to
investigate the evolution process of ACS under suppression. In this
model, we observe a discontinuous phase transition where a
microscopic independent ACS appears abruptly. The increase of its
size is found to grow linearly with the system size by simulations.
We derive the threshold $t_{\alpha}$ analytically and verify our
result through numerical simulations on different system sizes and
various choices of $m$. As the suppression increases, the phase
transition is dramatically deferred. Furthermore, we explore the
evolution process of the largest independent ACS. To quantify the
tolerance of the system to the emergence of a microscopic
independent ACS, we define a quantity $n_{\beta}$ that describes the
average number of largest independent ACSs during the evolution
process. It is shown $n_{\beta}$ increases as the selection rule
becomes more strict. Therefore, on average, a system with larger $m$
would contain more largest independent ACSs during the evolution.
Our model gives a possible explanation for the sudden appearance of
a class of species or chemical organizations in specific situations.
By only introducing a selection rule, the evolution of ACS presents
qualitative difference with that of the classical model. Our study
sheds light on the research of evolutional process of ACS and
provides helpful instructions to design effective strategies to
control the appearance of ACS in practice.

\section{Acknowledgements}

This work is supported by the National Natural Science Foundation of China No. 11290141 and No. 11201019.


\begin{thebibliography}{99}

\bibitem{Albert2002}
R. Albert and A.-L. Barab\'{a}si, Statistical mechanics of complex
networks, {\it Rev. Mod. Phys.} {\bf 74}, 47-97 (2002).

\bibitem{Fell2000}
D.A. Fell and A. Wagner, The small-world of metabolism, {\it Nature
Biotechnology} {\bf 18}, 1121-1122 (2000).

\bibitem{Jeong2000}
H. Jeong, B. Tombor, A. Albert, Z.N. Oltvai, and A.-L. Barab\'{a}si,
The large-scale organization of metabolic networks, {\it Nature}
{\bf 407}, 651-654 (2000).

\bibitem{Williams2000}
R.J. Williams and N.D. Martinez,Simple rules yield complex food
webs, {\it Nature} {\bf 404}, 180-183 (2000).

\bibitem{Camacho2002}
J. Camacho, R. Guimer\`{a}, and L.A.N. Amaral, Robust patterns in
food web structure, {\it Phys. Rev. Lett.} {\bf 88}, 228102 (2002).


\bibitem{Watts1998}
D.J. Watts and S.H. Strogatz, Collective dynamics of 'small-world'
networks, {\it Nature} {\bf 393}, 440-442 (1998).

\bibitem{Kauffman1986}
S.A. Kauffman, Autocatalytic sets of proteins, {\it J. Theor. Biol.}
{\bf 119}, 1-24 (1986).

\bibitem{Farmer1986}
J.D. Farmer, S.A. Kauffman, and N.H. Packard, Autocatalytic
replication of polymers, {\it Physica D: Nonlinear Phenomena} {\bf
22}, 50-67 (1986).

\bibitem{Fontana1994}
W. Fontana and L.W. Buss, "The arrival of the fittest": Toward a
theory of biological organization, {\it Bull. Math. Biol.} {\bf 56},
1-64 (1994).

\bibitem{Stadler1994}
P.F.Stadler, W. Fontana, and J.H. Miller, Random catalytic reaction
networks, {\it Physica D: Nonlinear Phenomena} {\bf 63}, 378-392
(1993).

\bibitem{Joyce1989}
G.F. Joyce, RNA evolution and the origins of life, {\it Nature} {\bf
338}, 217-223 (1989).

\bibitem{Bak1993}
P. Bak and K. Sneppen, Punctuated equilibrium and criticality in a
simple model of evolution, {\it Phys. Rev. Lett.} {\bf 71},
4083-4086 (1993).

\bibitem{Jain1998}
S. Jain and S. Krishna, Autocatalytic Sets and the Growth of
Complexity in an Evolutionary Model, {\it Phys. Rev. Lett.} {\bf
81}, 5684-5687 (1998).

\bibitem{Jain2002}
S. Jain and S. Krishna, Crashes, recoveries, and ¡°core shifts¡± in
a model of evolving networks, {\it Phys. Rev. E} {\bf 65}, 026103
(2002).

\bibitem{Eigen1971}
M. Eigen, Selforganization of matter and the evolution of biological
macromolecules, {\it Naturwissenschaften} {\bf 58}, 465-523 (1971).

\bibitem{Kauffman1971}
S.A. Kauffman, Celluar homeostasis, epigenesis and replication in
randomly aggregated macromolecular systems, {\it J. Cybernetics}
{\bf 1}, 71-96 (1971).

\bibitem{Rossler1971}
O.E. Rossler, A system theoretic model of biogenesis, {\it Z.
Naturforschung} {\bf 26b}, 741-746 (1971).

\bibitem{Jain2001}
S. Jain and S. Krishna, A model for the emergence of cooperation,
interdependence and structure in evolving networks, {\it Proc. Natl.
Acad. Sci. U.S.A.} {\bf 98}, 543-547 (2001).

\bibitem{Stauffer1994}
D. Stauffer and A. Aharony, {\it Introduction to Percolation Theory}
(Taylor \& Francis, London, ed. 2, 1994).

\bibitem{Erdos1960}
P. Erd\"{o}s and A. R\'{e}nyi, On the evolution of random graphs,
{\it Publ. Math. Inst. Hungar. Acad. Sci} {\bf 5}, 17 (1960).

\bibitem{Achlioptas2009}
D. Achlioptas, R.M. D'Souza, and J. Spencer, Explosive Percolation
in Random Networks, {\it Science} {\bf 323}, 1453 (2009).

\bibitem{Ziff2009}
R.M. Ziff, Explosive Growth in Biased Dynamic Percolation on
Two-Dimensional Regular Lattice Networks, {\it Phys. Rev. Lett.}
{\bf 103}, 045701 (2009).



\bibitem{Radicchi2009}
F. Radicchi and S. Fortunato, Explosive Percolation in Scale-Free
Networks, {\it Phys. Rev. Lett.} {\bf 103}, 168701 (2009).

\bibitem{Araujo2010}
N.A.M. Ara\'{u}jo and H.J. Herrmann, Explosive Percolation via
Control of the Largest Cluster, {\it Phys. Rev. Lett.} {\bf 105},
035701 (2010).

\bibitem{Cho2013}
Y.S. Cho, S. Hwang, H.J. Herrmann, and B. Kahng, Avoiding a Spanning
Cluster in Percolation Models, {\it Science} {\bf 339}, 1185-1187
(2013)

\bibitem{Chen2011}
W. Chen and R.M. D'Souza, Explosive Percolation with Multiple Giant
Components, {\it Phys. Rev. Lett.} {\bf 106}, 115701 (2011).

\bibitem{Zhang2013}
R. Zhang, W. Wei, B. Guo, Y. Zhang, and Z. Zheng, Analysis on the
evolution process of BFW-like model with discontinuous percolation
of multiple giant components, {\it Physica A} {\bf 392}, 1232-1245
(2013)

\bibitem{Costa2010}
R.A. da Costa, S.N. Dorogovtsev, A.V. Goltsev, and J.F.F. Mendes,
Explosive Percolation Transition is Actually Continuous, {\it Phys.
Rev. Lett.} {\bf 105}, 255701 (2010).

\bibitem{Riordan2011}
O. Riordan and L. Warnke, Explosive Percolation Is Continuous, {\it
Science} {\bf 333}, 322-324 (2011).

\bibitem{Grassberger2011}
P. Grassberger, C. Christensen, G. Bizhani, S. Son, and M.
Paczuski,Explosive Percolation is Continuous, but with Unusual
Finite Size Behavior, {\it Phys. Rev. Lett.} {\bf 106}, 255701
(2011).

\bibitem{Lee2011}
H.K. Lee, B.J. Kim, and H. Park, Continuity of the explosive
percolation transition, {\it Phys. Rev. E} {\bf 84}, 020101(R)
(2011).

\bibitem{Bastas2011}
N. Bastas, K. Kosmidis, and P. Argyrakis, Explosive site percolation
and finite-size hysteresis, {\it Phys. Rev. E} {\bf 84}, 066112
(2011).

\bibitem{Bollobas2001}
B. Bollobas, {\it Random Graphs}, 2nd ed., Academic Press, New York, (2011).


\end{thebibliography}
\end{document}